# Title: Explaining Recurring Maser Flares in the ISM Through Large-scale Entangled Quantum Mechanical States

**Authors:** Fereshteh Rajabi[1], Martin Houde[1,2]*

**Affiliations:**

[1]University of Western Ontario, Physics and Astronomy, London, ON, Canada

[2]California Institute of Technology, Division of Physics, Mathematics and Astronomy, Pasadena, California, USA

*Correspondence to: mhoude2@uwo.ca

**Abstract**: We apply Dicke's theory of superradiance introduced in 1954 to the methanol 6.7 GHz and water 22 GHz spectral lines, often detected in molecular clouds as signposts for the early stages of the star formation process. We suggest that superradiance, characterized by burst-like features taking place over a wide range of time-scales, may provide a natural explanation for the recent observations of periodic and seemingly alternating methanol and water maser flares in G107.298+5.639. Although these observations would be very difficult to explain within the context of maser theory, we show that these flares may result from simultaneously initiated 6.7-GHz methanol and 22-GHz water superradiant bursts operating on different time-scales, thus providing a natural mechanism for their observed durations and time ordering. The evidence of superradiance in this source further suggests the existence of entangled quantum mechanical states, involving a very large number of molecules, over distances up to a few kilometres in the interstellar medium.

**One Sentence Summary:** Simultaneously recurring maser flares from different molecular species/lines in the ISM can be explained with Dicke's superradiance.

**Main Text:** Since their first detection in the OH 18 cm lines (*1*), a large number of masers from several molecules were discovered in both galactic and extragalactic environments. The main characteristics of masers include high brightness temperatures, corresponding to very high emission intensities over small spatial scales, narrow line-widths, and occasionally high levels of polarization across the spectral lines (*2, 3*). These attributes of the maser action result from the stimulated emission process in a medium where a population inversion is established and maintained, leading to large amplifications along optical paths exhibiting good velocity coherence for the spectral line under consideration.

In addition, observations show that some maser sources exhibit significant intensity variability on time-scales ranging from days to several years. The 22-GHz water masers in Orion KL, for example, exhibited drastic flux density variations over a six-year period between 1979 and 1985 (*4, 5*). This phase of activity was followed by a twelve-year quiescent period that ended in 1997, when subsequent burst activity was detected in this source (*6*). Although the majority of flaring sources display abrupt changes of flux density through isolated impulsive phases (*7*), interestingly, intensity variations in some sources are sometimes found to be periodic, where the corresponding maser transition regularly alternates between phases of high activity and quiescence, (*8, 9*). Although a number of models are proposed to explain such time variations, the underlying mechanism for most of these observations still remains obscure (*10,11*).

Some of the aforementioned requirements for the maser action, i.e., population inversion and velocity coherence, are also necessary for superradiance; a fundamentally different radiation enhancement process. Superradiance, introduced by Dicke in 1954, is a coherent and cooperative quantum mechanical phenomenon by which a group of *N* inverted atoms/molecules emit a radiation pulse (burst) of intensity proportional to $N^2$. Although virtually unknown to astrophysicists, superradiance has become a very intense research field within the physics community since its introduction by Dicke (*12, 13*) and its initial laboratory confirmation by Skribanowitz et al. (*14*) (see also (*15, 16*)). As superradiant pulses can exhibit a temporal behavior resembling that of flares discovered for some masers in the circumstellar envelope (CSE) of evolved stars or elsewhere in the interstellar medium (ISM), we recently started investigating the possibility of superradiance within the context of astrophysics. We concluded that it could, in principle, take place in some regions when the necessary conditions are met (i.e., population inversion, velocity coherence, and long dephasing time-scales compared to those characterizing superradiance (*17, 18*)).

In this paper, we extend our analysis to the methanol 6.7-GHz and water 22-GHz maser transitions using the one-dimensional superradiance formalism presented in Rajabi & Houde (*17, 18*). We start with a brief summary of the important parameters used in our numerical analyses of recent observations of periodic and seemingly alternating flares of 6.7-GHz methanol and 22-GHz water masers in the G107.298+5.639 star-forming region. We end with a brief conclusion, while further observational evidence for superradiance in star-forming regions for the methanol (in G33.64-0.21) and water (in Cep A) lines is provided as Supplementary Material.

**Superradiance Model.** As was recently discussed in Rajabi & Houde (*17, 18*), a rapid and significant increase (sometimes followed by oscillations) in radiation intensity is a behavior typical of a (large-sample) superradiant system. In particular, the analyses of Rajabi & Houde (*17*) established that superradiance may provide a viable explanation for the observed OH 1612-MHz intensity bursts detected in the Mira star U Orionis (*19*) and the pre-planetary nebula IRAS 18276-1431 (*20*). The response of a superradiant system is characterized by a few parameters, the most important being the characteristic time-scale of superradiance $T_\mathrm{R}$, which is given by

$$T_\mathrm{R} = \tau_\mathrm{sp} \frac{8\pi}{3nL\lambda^2}, \tag{1}$$

where $\tau_\mathrm{sp}$ is the spontaneous decay time-scale of a single molecule (i.e., the inverse of the Einstein spontaneous emission coefficient), *n* the density of inverted molecules taking part in the superradiant process, *L* the length of the (cylindrical) large-sample (*nL* is thus the column density of inverted molecules), and $\lambda$ the wavelength of the radiation interacting with the molecules in the superradiance system. For a given transition, $\lambda$ and $\tau_\mathrm{sp}$ are fixed and $T_\mathrm{R}$ thus only depends on the column density of molecules partaking in superradiance.

When a superradiant system is inverted through some pumping mechanism and a critical threshold for the inverted column density is met or exceeded (see equation (3) below), the energy stored in the system is released after the so-called delay time $\tau_\mathrm{D}$ given by

$$\tau_D \simeq \frac{T_R}{4}\left|\ln\left(\frac{\theta_0}{2\pi}\right)\right|^2, \qquad (2)$$

with the initial Bloch angle $\theta_0 = 2/\sqrt{N}$, where $N$ is the number of inverted molecules in the sample (*15, 16*). While the delay time $\tau_D$ gives an estimate of the time when the first superradiance burst takes place, the characteristic time $T_R$ sets the duration of each burst. In a superradiance large-sample, the energy can be radiated away in a series of bursts, through a phenomenon known as the ringing effect. The number of ringing oscillations varies as a function of $T'$, the time-scale of the most important dephasing effect (e.g., collisions) that will tend to work against the superradiance phenomenon.

In a more general sense, superradiance can be only observed if $\tau_D < T'$ (see (*17, 18*) for more details), and this implies the existence of the aforementioned threshold in inverted-population column density

$$(nL)_{\text{crit}} \approx \frac{2\pi}{3\lambda^2}\frac{\tau_{\text{sp}}}{T'}\left|\ln\left(\frac{\theta_0}{2\pi}\right)\right|^2 \qquad (3)$$

that must be met for the initiation of superradiance. More precisely, for column densities below this critical value the dephasing effects prevent coherent interactions and the system operates in a maser regime. As soon as $nL \geq (nL)_{\text{crit}}$ the system switches to a superradiance mode and the masing region breaks into a large number of superradiance large-samples for which this condition is met (*17*). Whether $(nL)_{\text{crit}}$ is crossed through a slow increase in pumping of the inverted column density or a fast population-inverting pulse is irrelevant. It only matters that the condition $nL \geq (nL)_{\text{crit}}$ is somehow reached.

In order to test our superradiance model, we have chosen $T_R$ and $T'$ as free parameters in the fitting process to intensity curves given by the data presented below. It must however be noted that, as was observed by Rajabi & Houde (*17*) for the OH 1612 MHz line superradiance bursts in U Orionis and IRAS 18276-1431, the volume occupied by a single superradiant large-sample is several orders of magnitude smaller than a typical maser region. A similar statement applies to the cases studied in this paper. It follows that the superradiance intensity curves must result from the contributions of a very large number of separate but approximately simultaneously triggered superradiant samples. We have therefore augmented our one-dimensional superradiance model (*17*) to account for this by averaging over several realizations of superradiance samples for which a common $T'$ is used. These realizations result from a Gaussian-distributed ensemble of $T_R$ values of mean $\langle T_R \rangle$ and standard deviation $\sigma_{T_R}$.

**Methanol 6.7-GHz and Water 22-GHz Flares in G107.298+5.639.** G107.298+5.639 is an intermediate-mass young stellar object deeply embedded in the molecular cloud L1204/S140 (*21*) at a distance of ~0.9 kpc (*22*). For the periods of July to December 2014 and similarly in 2015, Szymczak et al. (*21*) monitored methanol 6.7-GHz and water 22-GHz masers in this source with the Torun 32-m Radio Telescope. During high activity intervals, the methanol observation rate was increased to eight times a day whereas in quiescent periods they were conducted only once a week. The water observations were, however, repeated daily with eight gaps of 4 to 5 days of no observations.

The observations of the methanol 6.7-GHz masers indicated four spectral features, among which the $v_{lsr} = -7.4$ km s$^{-1}$, $-8.6$ km s$^{-1}$, and $-9.2$ km s$^{-1}$ components exhibited a 34.4-day cyclic behavior. Faint cyclic emission was also detected in a few other features in the velocity range of $v_{lsr} = -17.2$ km s$^{-1}$ to $-13.6$ km s$^{-1}$, although these components were not visible in all cycles. The methanol components showed strong flux variations in the form of repeating (lone) bursts lasting for four to twelve days.

The 22 GHz water maser emission was also detected in six spectral features at velocities ranging from $v_{lsr} = -18.3$ km s$^{-1}$ to $-1.1$ km s$^{-1}$. Importantly, some features (e.g., those near $v_{lsr} = -16.5$ km s$^{-1}$ and $-8$ km s$^{-1}$) peaked in intensity with the same periodicity as, but delayed relative to, methanol flares at velocities located within $\pm 1.1$ km s$^{-1}$ of those of the water masers themselves. A similar behavior was detected at a velocity near $v_{lsr} = -11.0$ km s$^{-1}$ where methanol and water flares were also observed in alternation. Based on high angular resolution data sets collected with the European VLBI Network (EVN) and VLBI Exploration of Radio Astrometry (VERA) (*23*), Szymczak et al. (*21*) concluded that some of the periodically alternating methanol and water flares originate from the same molecular gas volume of 30 to 80 AU in size.

In order to find a viable explanation for these observations a number of scenarios were examined, but none were able to adequately reproduce the observations. The alternation of the water and methanol maser bursts is the main feature to be explained. However, this is very difficult to achieve within the context of maser theory, even with the assumptions that the two types of masers happen in the same region and are being periodically enhanced by some pulsating pumping source. For example, it is hard to conceive how the water and methanol maser features would then occur in alternation while also showing different time durations for their flares. We now show how this kind of behavior could naturally arise, and may be expected, when studied within the context of Dicke's superradiance.

We assume that we are in the presence of a periodically changing pumping source that simultaneously acts on the population levels of both the 22 GHz water and 6.7 GHz methanol transitions. Although we are not aware of any other observations (beside the periodic maser flares discussed here) that could provide evidence for such a scenario in G107.298+5.639, we know of at least one other young protostellar system where strong, cyclic variations in infrared luminosity have been observed with a period comparable to that seen in G107.298+5.639 (i.e., 25.34 day for LRLL 54361 (*24*)). Such infrared intensity variations could, in principle, directly affect the pumping level of maser transitions.

It is important to note that for a given inverted column density the value of $T_R$, and therefore the duration of superradiant bursts for a spectral line scales as $\tau_{sp}/\lambda^2$ (see equation (1) ). It follows that under similar conditions (i.e., assuming for the moment $(nL)_{CH_3OH} \approx (nL)_{H_2O}$) we should expect a superradiance time-scale ratio of approximately $1:8.7$ between the methanol 6.7 GHz and water 22 GHz lines, respectively. In fact, this expected relationship between $T_{R,CH_3OH}$ and $T_{R,H_2O}$ provides us with the needed element to explain the observations of Szymczak et al.

(*21*). That is, since $T_\mathrm{R}$ sets both the duration of a superradiant burst and the time delay $\tau_\mathrm{D}$ before its emergence (see equation (2)) it is to be expected that the methanol flares will be narrower and appear earlier than those of water. Evidently, it is unlikely that the inverted column densities for methanol at 6.7 GHz and water at 22 GHz will be the same, and we thus relax this approximation in what follows. But the above scenario should still hold for cases where their difference is not too pronounced.

Given the observed duration of flares for the methanol 6.7-GHz and water 22-GHz masers, we adjusted the values of $T_\mathrm{R}$ for these transitions to reproduce superradiant bursts of similar time-scales (i.e., approximately 10 days for methanol and 30 days for water). The result of our analysis is shown in Figure 1. The superradiance intensity models (solid curves) were calculated using ensembles of 1000 superradiance large-samples (the shape of an intensity curve converges after a few hundred realizations are averaged) tailored to the flaring event occurring between MJD 57,260 to 57,300 in Figure 3 of Szymczak et al. (*21*) (i.e., Day 7260 to 7300 in our Figure 1) for the $v_\mathrm{lsr} = -8.57$ km s$^{-1}$ 6.7-GHz methanol (red dots) and $v_\mathrm{lsr} = -7.86$ km s$^{-1}$ 22-GHz water (blue dots) spectral features. As mentioned earlier, such flares repeated on an approximately 34.4-day period (*21*). For methanol the model parameters are $\langle T_\mathrm{R} \rangle_{\mathrm{CH_3OH}} = 2.1$ hr, $\sigma_{T_\mathrm{R,CH_3OH}} = 0.07 \langle T_\mathrm{R} \rangle_{\mathrm{CH_3OH}}$, and $T'_{\mathrm{CH_3OH}} = 90 \langle T_\mathrm{R} \rangle_{\mathrm{CH_3OH}}$, yielding a mean inverted column density of $\langle nL \rangle_{\mathrm{CH_3OH}} \approx 3.5 \times 10^4$ cm$^{-2}$, while for water we have $\langle T_\mathrm{R} \rangle_{\mathrm{H_2O}} = 7.7$ hr, $\sigma_{T_\mathrm{R,H_2O}} = 0.04 \langle T_\mathrm{R} \rangle_{\mathrm{H_2O}}$, $T'_{\mathrm{H_2O}} = 70 \langle T_\mathrm{R} \rangle_{\mathrm{H_2O}}$, and $\langle nL \rangle_{\mathrm{H_2O}} \approx 8.4 \times 10^4$ cm$^{-2}$. In both cases, the models were scaled in intensity to the data. As seen in the figure the methanol superradiance curve provides a very good fit to the corresponding data, especially in the wings, while although there is a fair amount of scatter in the data, the water superradiance intensity curve captures well the overall behavior of the water flare. It is important to note that, *despite the apparent time ordering in the emergence of the methanol and water flares, both superradiance models were initiated at the same time* on Day 7261.5. The alternation between the methanol and water bursts observed in Figure 3 of Szymczak et al. (*21*) may thus be readily, and simply, explained by the fact that $\langle T_\mathrm{R} \rangle_{\mathrm{H_2O}} \approx 3.7 \langle T_\mathrm{R} \rangle_{\mathrm{CH_3OH}}$, which would delay the appearance of the water flare (from equation (2)) and broaden it relative to methanol. We note that strict simultaneity in the excitation of the two species is not an absolute requirement for superradiance to fit the data, but it is telling that it can provide a viable model even if simultaneity is realized. Although one could conceive of models based on maser theory alone to account for the different time scales between the water and methanol intensity bursts (e.g., by invoking non-radiative excitation processes such as grain mantle evaporation or dust heating, followed by re-emission), our model has the advantage of being simpler as it is solely based on the difference between the characteristic time scales $\langle T_\mathrm{R} \rangle_{\mathrm{H_2O}}$ and $\langle T_\mathrm{R} \rangle_{\mathrm{CH_3OH}}$. The two inverted-population column densities have comparable mean values (i.e., $\langle nL \rangle_{\mathrm{H_2O}} \approx 2.4 \langle nL \rangle_{\mathrm{CH_3OH}}$) and correspond to large-samples lengths of $\langle L \rangle_{\mathrm{CH_3OH}} \approx 3.5 \times 10^5$ cm and $\langle L \rangle_{\mathrm{H_2O}} \approx 8.4 \times 10^4$ cm when $\langle n \rangle_{\mathrm{CH_3OH}} = 0.1$ cm$^{-3}$ and $\langle n \rangle_{\mathrm{H_2O}} = 1$ cm$^{-3}$. As stated earlier, these length scales are markedly smaller than those typical of masers, implying the presence of a large number of superradiance samples.

Finally, it remains to be explained how the broader water intensity flares stay synchronized to the overall 34.4-day period between successive events, as shown in Figure 3 of Szymczak et al. (*21*). That is, the methanol bursts are short and end well before the appearance of the subsequent flaring event, but our water superradiance model in Figure 1 seems to indicate that the tail of the corresponding curve extends beyond the start of the next eruption cycle. The observed synchronization probably results from the fact that the arrival of the new pumping event responsible for the next flares resets the large-samples by re-establishing the inverted populations, which in effect terminates the water superradiance cascade and associated flare (i.e., truncates our superradiance model curve). The intensity then tends to zero but superradiance is once more triggered, resulting in the appearance of the next burst in intensity. This regenerative pumping phenomenon has previously been observed in laboratory superradiance experiments (*25, 26*).

**Conclusion.** When combined with the recently reported evidence for this phenomenon in the environments of evolved stars (*17*), the discovery of superradiance in star-forming regions would broaden the applicability of our model and further establish the existence of a previously unsuspected physical phenomenon in the ISM. The occurrence of superradiance in astrophysical objects would also imply the presence of entangled quantum mechanical systems, involving a very large number of molecules, over distances up to a few kilometres in the ISM, which can also be of interest for quantum information research.

**Materials and Methods.** The G107.298+5.639 water 22 GHz and methanol 6.7 GHz data discussed in the main part of the paper were published by Szymczak et al. (*21*) and provided for our analysis by the authors. The G33.64-0.21 methanol 6.7 GHz data presented in the Supplementary Materials were previously published by Fujisawa et al. (*28*) and provided to us by the authors, while the Cepheus A water 22 GHz observations were taken from Mattila et al. (*7*). All these data sets were analyzed by numerically solving the one-dimensional Sine-Gordon equation (*15, 16, 17, 18*). As stated earlier, we augmented this model by averaging over several realizations of cylindrical superradiance samples taken from a Gaussian-distributed ensemble of $T_\mathrm{R}$ values of mean $\langle T_\mathrm{R} \rangle$ and standard deviation $\sigma_{T_\mathrm{R}}$.

In order to minimize diffraction and transverse effects not included in our one-dimensional model, we set the dimensions of the superradiance samples by imposing a Fresnel number of unity (*15*). This has also for effect to ensure that the size of the samples does not exceed the condition necessary for phase coherence to be maintained across their length (i.e., $L \sim \lambda/\phi_\mathrm{B}$ with $\lambda$ the wavelength and $\phi_\mathrm{B}$ the beam solid angle of the radiation, see for example (*3*)).

**Acknowledgments:** We are grateful to M. Szymczak for sharing her data of G107.298+5.639. We would also like to thank K. Fujisawa for making his data of G33.64-0.21, provided as Supplementary Material, available to us. We are grateful to Profs. S. Metchev and J. Cami for their useful suggestions and comments. **Funding:** M.H.'s research is funded through the NSERC Discovery Grant RGPIN-2016-04460 and the Western Strategic Support for Research Accelerator Success. **Author Contributions:** Both authors contributed to the science discussion and to the writing of the paper. F.R. obtained and analyzed the data and M.H. produced the figures appearing in the paper. **Competing interests:** The authors declare that they have no competing interests. **Data and materials availability:** All data needed to evaluate the conclusions in the paper are present in the paper and/or the Supplementary Materials. Additional data related to this paper may be requested from the authors.

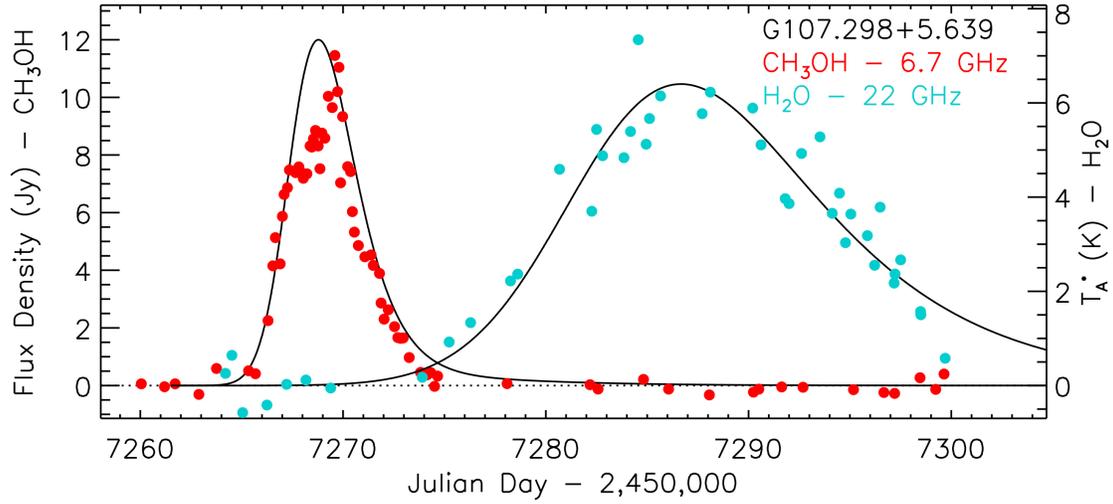

**Fig. 1.** Superradiance models (solid curves) for the G107.298+5.639 flaring event occurring between MJD 57,260 to 57,300 in Figure 3 of Szymczak et al. (*21*) (i.e., Day 7260 to 7300 in our figure) for the $v_{lsr} = -8.57$ km s$^{-1}$ 6.7-GHz methanol (red dots) and $v_{lsr} = -7.86$ km s$^{-1}$ 22-GHz water (blue dots) spectral features. Such flares repeated on an approximately 34.4-day period (*21*). The methanol and water data/models use the vertical axes on the left- and right-hand sides, respectively. For methanol the model parameters are $\langle T_R \rangle_{CH_3OH} = 2.1$ hr, $\sigma_{T_{R,CH_3OH}} = 0.07 \langle T_R \rangle_{CH_3OH}$, and $T'_{CH_3OH} = 90 \langle T_R \rangle_{CH_3OH}$, yielding a mean inverted column density of $\langle nL \rangle_{CH_3OH} \approx 3.5 \times 10^4$ cm$^{-2}$, while for water we have $\langle T_R \rangle_{H_2O} = 7.7$ hr, $\sigma_{T_{R,H_2O}} = 0.04 \langle T_R \rangle_{H_2O}$, $T'_{H_2O} = 70 \langle T_R \rangle_{H_2O}$, and $\langle nL \rangle_{H_2O} \approx 8.4 \times 10^4$ cm$^{-2}$. Both superradiance models were initiated at the same time, on Day 7261.5.

**Supplementary Materials:**

Supplementary text

Figures S1-S2

References (*27-30*)

# Supplementary Materials for

Explaining Recurring Maser Flares in the ISM Through Large-scale Entangled Quantum Mechanical States

Fereshteh Rajabi[1], Martin Houde[1,2]*

correspondence to: mhoude2@uwo.ca

**This PDF file includes:**

    Supplementary Text
    Figs. S1 to S2

**Supplementary Text**

Methanol 6.7 GHz Flares in G33.64-0.21.

G33.64-0.21 is a high-mass star-forming region located at a kinematic distance of 4.0 kpc with an estimated infrared luminosity of $1.2 \times 10^4 L_\odot$ (*27*). The spectra of the 6.7 GHz methanol masers in G33.64-0.21 were monitored daily with the Yamaguchi 32-m Radio Telescope over several time intervals from 2009 to 2015 by Fujisawa et al. (*28*). The corresponding observations identified five narrow maser spectral features (Components I to V, defined with increasing line-of-sight velocity; see Figure 1 of Fujisawa et al. (*28*)) with line widths of approximately $0.3 \, \text{km s}^{-1}$. Two bursts of radiation, lasting on the order of ten days, were observed in Component II ($v_\text{lsr} = 59.6 \, \text{km s}^{-1}$) in July and August 2009, while all other velocity components did not exhibit any significant change in their flux densities over similar time-scales or longer (see Figure 2 of Fujisawa et al. (*28*)). During both events, the flux densities increased approximately sevenfold within 24 hours and then returned to their original value while exhibiting a damped oscillator behavior. Subsequent observations with the Japanese VLBI Network revealed that Component II, responsible for the two bursts, emanates from the southwestern edge of G33.64-0.21 within a region measured to be much smaller than 70 AU. Different scenarios were proposed to explain these observations, but none were so far able to adequately describe an energy release mechanism responsible for such bursting behavior.

Given the damped oscillator character of the intensity curve during the bursts, we investigated the possibility of superradiance in the 6.7 GHz methanol line in an attempt to explain the energy relaxation mechanism at play for G33.64-0.21. Here, we focus on the second burst appearing in August 2009 in Figure 2 of Fujisawa et al. (*28*). The results of our analyses show that a group of methanol superradiance large-samples of mean inverted column density $\langle nL \rangle \sim 7 \times 10^4 \, \text{cm}^{-2}$ (e.g., of density $\langle n \rangle \sim 0.1 \, \text{cm}^{-3}$ and length $\langle L \rangle \sim 10^6 \, \text{cm}$) can reproduce similar intensity variations as that of the 6.7 GHz line detected in G33.64-0.21. In Figure S1 we show the average intensity (scaled to the data) obtained with 1000 such superradiance large-samples calculated using our one-dimensional model (solid blue curve) superposed on the data from Fujisawa et al. (*28*) (black dots). The superradiance sample realizations are generated using $\langle T_\text{R} \rangle = 1.1 \, \text{hr}$, $\sigma_{T_\text{R}} = 0.07 \langle T_\text{R} \rangle$ and $T' = 600 \langle T_\text{R} \rangle$. As seen in the figure, our superradiance model agrees well with the data and is successful in reproducing the main characteristics of the observed intensity curve.

The observations of Fujisawa et al. (*28*) were initially carried out daily (from Day 5039 to Day 5043 in Figure S1) followed by alternate day monitoring of the source. As a result, the data are sparse considering the rapid intensity variations exhibited by the superradiance curve. This also implies that the peak flux density detected by Fujisawa et al. (*28*) may not represent the actual maximum experienced by the source; our model indicates a peak flux density of 350 Jy late on Day 5042. Finally, the dephasing time-scale $T'$ used to produce the solid curve is on the order of a month, which is reasonable within the expected gas densities $10^4 \, \text{cm}^{-3} < n_{\text{H}_2} < 10^9 \, \text{cm}^{-3}$ and temperature $T < 300 \, \text{K}$ in G33.64-0.21 (*29*).

Water 22 GHz Flares in Cepheus A

Cepheus A (Cep A) with 14 compact HII regions is a high-mass star-forming site located at a distance of $\sim 0.7$ kpc (*7*). In 1978, the 22-GHz water observations toward this source revealed significant time variability to be followed by a strong burst at $v_{lsr} = -8 \, \text{km s}^{-1}$ between April and December 1980. In October 1980, when this burst was in its decay phase Mattila et al. (*7*) started a three-year monitoring program of Cep A using the 14-m radio telescope of the Metsähovi Radio Research Station. This source was monitored through monthly observations until October 1983, except for a few time-spans where observations were repeated daily or every few days. In April 1983, the flux density of the water 22-GHz line at $v_{lsr} = -11.2 \, \text{km s}^{-1}$ increased six fold over 10 days reaching its maximum value of 1700 Jy on April 18. Later on, over the following 40 days, the flux density decayed to a background value while exhibiting a damped oscillator behavior. During this phase, a few secondary maxima were detected every 15 days or so. Different models were used to reproduce the observed light curve for this burst, but they were either unsuccessful in replicating the time-scale of the event or did not capture the secondary maxima.

In Figure S2 we show a superradiance intensity curve (scaled to the data) calculated using an ensemble of 1000 superradiance large-samples (solid blue curve) superposed on data from Figure 4c of Mattila et al. (*7*) (black dots). The superradiance realizations are produced using $\langle T_R \rangle = 8.2 \, \text{hr}$, $\sigma_{T_R} = 0.1 \langle T_R \rangle$ and $T' = 700 \langle T_R \rangle$. As seen in the figure the superradiance curve (solid curve) occurs over similar time-scale as that for the data and analogously exhibits a peak followed by secondary maxima as it damps. The relative intensities of the secondary maxima match those of the data reasonably well while the main peak exceeds the data. Given the simplicity of our superradiance model we can conclude that the overall behaviour of the burst is well captured by this model.

It must be noted that the water rotational energy levels ($J_{K_aK_c} = 6_{16}$ and $5_{23}$) corresponding to the 22-GHz line are degenerate and, in principle, superradiance can simultaneously operate in more than one of the corresponding hyperfine components. This can complicate the line flux density analysis due to the variation of the relative intensity of the degenerate transitions and their time of occurrence, which our superradiance model does not account for since it employs a two-level system approximation. This may be partly responsible for the disagreement between the model's peak intensity and the data. Another factor that may have an impact is the large half-power beam width ($\approx 4'$) of the telescope used for these observations, which were inevitably sensitive to an extended region and perhaps suffer contamination from a number of sources.

Once again, the results of our analysis suggest that a large group of water superradiance samples of mean inverted column density $\langle nL \rangle \sim 6 \times 10^4 \, \text{cm}^{-2}$ (e.g., $\langle n \rangle \sim 1 \, \text{cm}^{-3}$ and $\langle L \rangle \sim 10^5 \, \text{cm}$) must be responsible for the observed radiation intensity. We also note that the dephasing time-scale $T' = 700 \langle T_R \rangle$ or 238 days resulting from our calculations is less restrictive than the estimated collision time-scales for a given molecular hydrogen density $n_{H_2} = 10^8 \, \text{cm}^{-3}$ at $T \sim 100 \, \text{K}$ to $\sim 200 \, \text{K}$ consistent with the pumping model of water masers (*21, 30*).

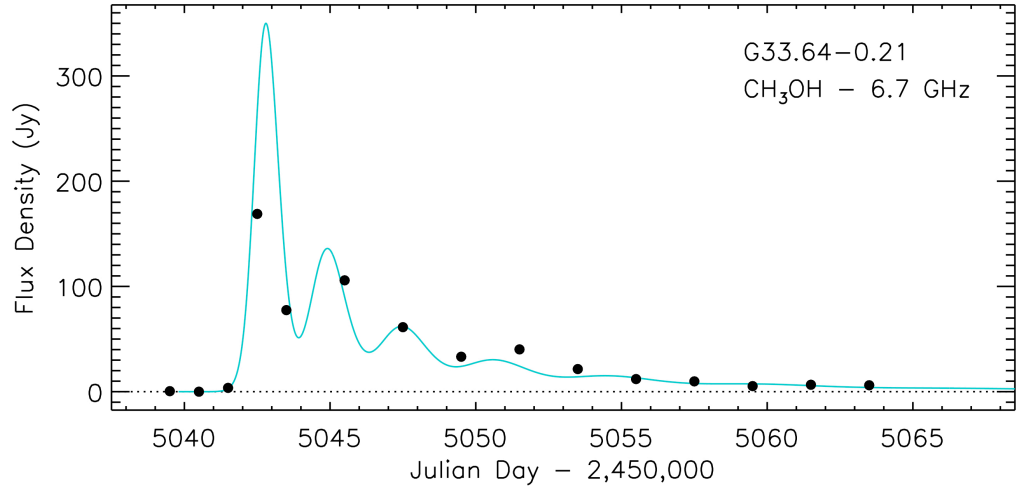

**Fig. S1.**

A superradiance large-sample intensity model (solid blue curve) superposed on the data Fujisawa et al. (*28*) (black dots) obtained in July and August 2009 for the second methanol 6.7-GHz burst in G33.64-0.21. The superradiance intensity is averaged over 1000 large-samples taken from a Gaussian-distributed ensemble of $T_R$ values of mean and standard deviation of $\langle T_R \rangle = 1.1\,\mathrm{hr}$ and $\sigma_{T_R} = 0.07 \langle T_R \rangle$, respectively, and scaled to the data. The dephasing time scale was set to $T' = 600 \langle T_R \rangle$ for all samples, and the superradiance pulses were initiated from internal fluctuations characterized by an initial Bloch angle $\langle \theta_0 \rangle \sim 10^{-6}\,\mathrm{rad}$.

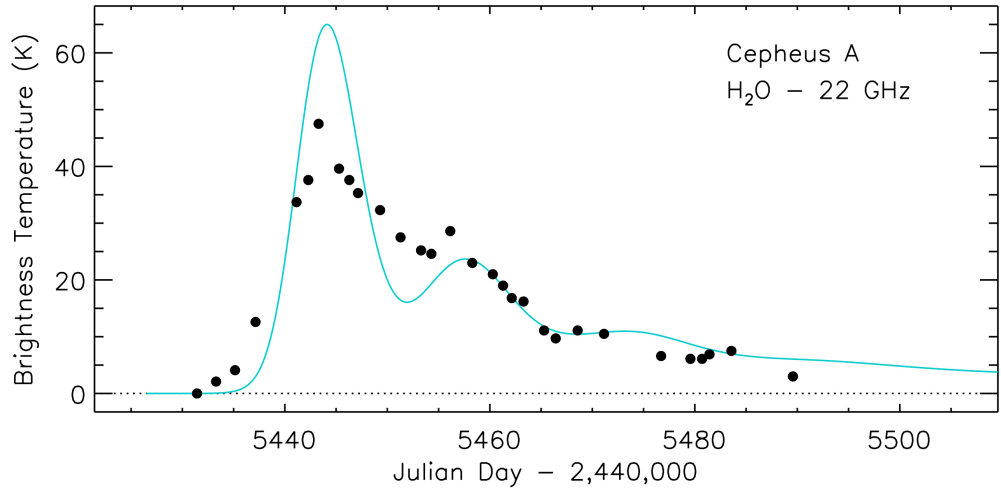

**Fig. S2**
A superradiance large-sample intensity model (solid blue curve) superposed on data from Mattila et al. (*7*) (black dots) obtained in April and May 1983 for the water 22-GHz burst at $v_{lsr} = -11.2\,\mathrm{km\,s^{-1}}$ in Cep A. The superradiance intensity is generated using 1000 superradiance realizations with $\langle T_R \rangle = 8.2\,\mathrm{hr}$, $\sigma_{T_R} = 0.1 \langle T_R \rangle$, and a dephasing time-scale $T' = 700 \langle T_R \rangle$.